# BUSCO update: novel and streamlined workflows along with broader and deeper phylogenetic coverage for scoring of eukaryotic, prokaryotic, and viral genomes

Mosè Manni*, Matthew R Berkeley*, Mathieu Seppey*, Felipe A Simão, and Evgeny M Zdobnov[+]

Department of Genetic Medicine and Development, University of Geneva, and Swiss Institute of Bioinformatics, Geneva, Switzerland

* contributed equally
[+]Corresponding author; email: Evgeny.Zdobnov@unige.ch

**Abstract**

Methods for evaluating the quality of genomic and metagenomic data are essential to aid genome assembly and to correctly interpret the results of subsequent analyses. BUSCO estimates the completeness and redundancy of processed genomic data based on universal single-copy orthologs. Here we present new functionalities and major improvements of the BUSCO software, as well as the renewal and expansion of the underlying datasets in sync with the OrthoDB v10 release. Among the major novelties, BUSCO now enables phylogenetic placement of the input sequence to automatically select the most appropriate dataset for the assessment, allowing the analysis of metagenome-assembled genomes of unknown origin. A newly-introduced genome workflow increases the efficiency and runtimes especially on large eukaryotic genomes. BUSCO is the only tool capable of assessing both eukaryotic and prokaryotic species, and can be applied to various data types, from genome assemblies and metagenomic bins, to transcriptomes and gene sets.

**Introduction**

Advances in sequencing technologies are powering accumulation of genomics data at an accelerating rate, from sequencing isolates and single cells to metagenomes of entire microbial communities. In turn, computational genomics approaches are essential to digest such molecular data into our evolving understanding of genome diversity across the tree of life, from microbes to animals and plants. Evaluating the quality of genomic data, in terms of completeness and redundancy, is critical for subsequent analyses and for the correct interpretation of the results. Complementing technical measures like the N50 value, biologically meaningful metrics based on expected gene content have proved to be useful for estimating the quality of genomes, as exemplified by our BUSCO tool (Simão et al. 2015; Waterhouse et al. 2018), the now discontinued CEGMA (Parra et al. 2007), CheckM (Parks et al. 2015) aimed at prokaryotes, EukCC (Saary et al. 2020) proposed for microbial eukaryotes, and CheckV (Nayfach et al. 2020) for viruses. The latest BUSCO versions introduce new functionalities for assessments of eukaryotic, prokaryotic and viral data, along with improvements in runtimes and user experience. The underlying datasets have been renewed and expanded in sync with the OrthoDB v10 release (Kriventseva et al. 2019; Zdobnov et al. 2021), providing coverage of many more lineages and a revised baseline with increased data sampling.

Here we describe the new functionalities and datasets introduced after the release of BUSCO v3. With respect to v3, the last BUSCO version, v5, features *i)* a major upgrade of the underlying datasets in sync with OrthoDB v10; *ii)* an updated workflow for the assessment of prokaryotic and viral genomes using the gene predictor Prodigal (Hyatt et al. 2010); *iii)* an alternative workflow for the assessment of eukaryotic genomes using the gene predictor MetaEuk (Levy Karin et al. 2020); *iv)* a workflow to automatically select the most appropriate BUSCO dataset, enabling the analysis of sequences of unknown origin; *v)* an option to run batch analysis of multiple inputs to facilitate high-throughput assessments of large data sets and metagenomic bins; *vi)* a major refactoring of the code, and maintenance of two distribution channels on Bioconda (Grüning et al. 2018) and Docker (Merkel 2014). These developments make BUSCO suitable for comprehensive analyses of large heterogeneous datasets, from large eukaryotic genomes to metagenome-assembled genomes (MAGs) of microbial eukaryotes, prokaryotes and viruses.

**Results and Discussion**
*Upgrade of datasets*

The benchmarking datasets of single-copy orthologs were revised and expanded using the v10 release of OrthoDB (www.orthodb.org), which provides evolutionary and functional annotations of orthologs among the most comprehensive sampling of genomic diversity. The creation of novel datasets was necessary to cover more lineages with higher-resolution datasets and to revise the existing ones with increased and evenly sampled data across the phylogenetic tree. Following the strategy devised previously (Simão et al. 2015; Waterhouse et al. 2018) and taking advantage of the greatly increased number of representative species, we compiled 193 odb10 (OrthoDB v10) datasets (table 1), more than a threefold increase over odb9 sets, and comprising overall a fivefold increase in the number of BUSCO marker genes (hereafter BUSCOs) derived from more than twice as many species as in the previous datasets version. Supplementary table 1 lists the available BUSCO odb10 datasets along with the number of markers and species used to construct the sets. BUSCO v5 also includes 27 viral datasets, supporting the analysis of a subset of viruses. We compared the estimates of completeness of BUSCO v5 with v3 for a self-validation of the major datasets on gene sets of Bacteria, Fungi and Metazoa (fig. 1 and supplementary table 2). Figure 1 shows a good concordance overall, especially for Metazoa. Slightly more conservative estimates of v5 over v3 for fungi can be explained by the higher number of markers in odb10 BUSCO datasets compared to odb9, and by a more than sixfold increase in the number of species. Figure 1b shows a cluster of fungal genomes with BUSCO v5 scores diverging from v3. These species belong to Microsporidia, a group of early diverging fungi once thought to be protozoans (Wadi and Reinke 2020). Their genomes are known to have a reduced set of genes that are commonly present in fungi and other eukaryotes. The discrepancy between v3 and v5 reflects the increased number of markers for the fungi_odb10 dataset which is more balanced towards the majority of fungal clades (supplementary fig. 1). The "microsporidia_odb10" panel in figure 2a displays the assessment of microsporidian genomes with the most specific dataset microsporidia_odb10, which yields a more accurate assessment of these genomes. This example highlights the importance of using the most specific BUSCO dataset available for the species of interest, as large differences in terms of gene content can often occur within higher taxonomic levels. Nevertheless, to obviate the biased estimation when the fungi_odb10 is used on microsporidians we also added a "parasitic check" that recalculates the scores based on the list of fungal markers missing in these species (see supplementary notes).

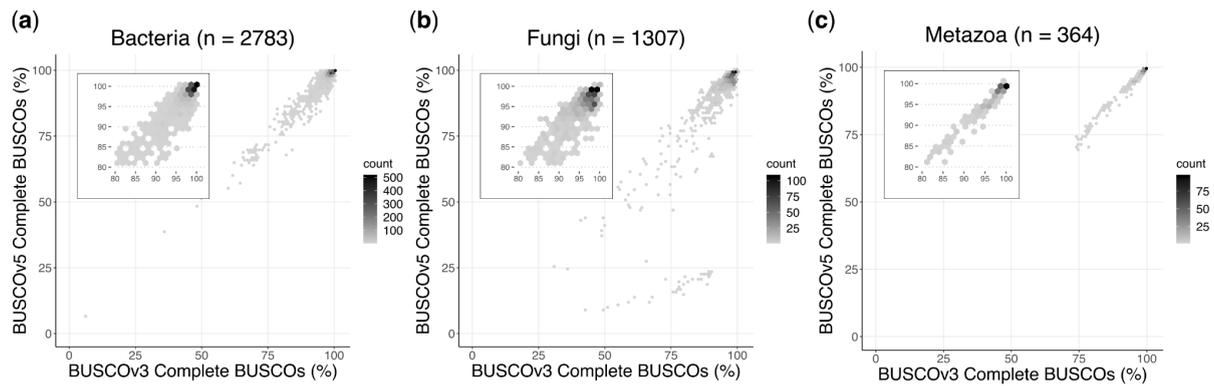

**Fig. 1.** Comparison of the number of complete BUSCOs obtained by running BUSCO v5 and v3 with BUSCO odb_10 and odb_9 datasets on (*a*) bacterial, (*b*) fungal and (*c*) metazoan gene sets.

*Novel BUSCO workflows*

The BUSCO software was revised and new functionalities introduced to enable the analysis of inputs of unknown taxonomic origin, and to improve assessments, throughput and runtimes. A breakdown of the features in v5 and the differences with v3 are described in the schema of supplementary figure 2, supplementary table 3 and supplementary notes.

*New prokaryotic and eukaryotic workflows.* The BUSCO software was revised to integrate the gene predictor Prodigal (Hyatt et al. 2010) for improving the analysis of prokaryotic genomes. The use of Prodigal coupled with the expanded number of datasets now available for Bacteria and Archaea (83 more datasets with respect to v3) make BUSCO a suitable choice when assessing prokaryotic data. A common issue when predicting genes is to select the correct genetic code (i.e. translation table) for the species under analysis. Each prokaryotic BUSCO dataset now contains information on the potential genetic codes characteristic of the species within the corresponding lineage. BUSCO selects the most likely genetic code automatically based on which code yields the highest coding density.

BUSCO v5 features a new workflow for the analysis of eukaryotic genomes that employs the gene predictor MetaEuk (Levy Karin et al. 2020), which relies on MMseqs2 (Steinegger and Söding 2017). MetaEuk was introduced to improve the assessment of large genomes for which the previous workflow was suffering from long runtimes. Two consecutive MetaEuk runs are implemented, and parameters are tuned differently for the second run to search for BUSCO genes missing after the first run. The BUSCO_MetaEuk workflow is the default option in v5, as it allows faster assessments. The Augustus gene predictor (Stanke et al. 2008) is still available in BUSCO and can be selected by specifying "--augustus" when running the analysis in "genome" mode. Since the two workflows use gene predictors that are based on different methods, it is expected to obtain non-identical results when assessing the same genome. Nevertheless, they produce comparable completeness estimations on genomic sequences, which at times outperform the completeness of corresponding gene sets (fig. 2a, supplementary fig. 3, supplementary table 4 and supplementary notes). Figure 2a compares the results of BUSCO v5 using the two gene predictors on fungal genomes and their corresponding gene sets. Comparisons on arthropod and protist data are reported in supplementary figures 3b and c. On a set of 139 arthropod genomes, we investigated if there is a set of BUSCOs consistently missed by one workflow and found by the other one. There were no consistent major differences in the ability to predict specific BUSCOs on this set, apart from a couple of exceptions (supplementary notes, supplementary fig. 4).

      Supplementary figure 3 shows the substantial improvement in runtimes (runtimes axis has a log10 scale) when using the BUSCO_MetaEuk workflow. The higher speed is especially useful when assessing large genomes. For example, BUSCO_MetaEuk runs in 8h and 50 min (using 56 CPUs) on the 10.7-Gbp genome of the wheat *Triticum dicoccoides* (accession: GCF_002162155.1) with the poales_odb10 dataset (BUSCO score: C:99.1% [S:9.6%, D:89.5%], F:0.1%, M:0.8%, n:4896), while the BUSCO_Augustus workflow takes several days to complete. Even faster analyses can be obtained by reducing the sensitivity value (-s) of the two runs, via the "--metaeuk_parameters" and "--metaeuk_rerun_parameters" options (default values are s=4.5 and s=6 for the first and second run, respectively) (fig. 2b). For example, the runtime on the *T. dicoccoides* genome decreases to 2h and 24 min using a sensitivity value of s=3 (BUSCO score: C:98.7% [S:9.6%, D:89.1%], F:0.2%, M:1.1%, n:4896). However, changing the sensitivity values can have an impact on the estimates (fig. 2c, supplementary table 5). The default values were chosen as a trade-off between accuracy and runtimes. In most settings it is not advisable to change the sensitivity values in order to keep BUSCO results comparable. Nevertheless, having this option can be convenient when assessing very large genomes or for getting faster evaluations on preliminary assemblies. Assembling genomes is an iterative process in which multiple draft assemblies are often produced to compare the outcome of different parameters/pipelines. Using smaller sensitivity values facilitates quick draft BUSCO assessments that were not feasible before, and should speed up the overall genome assembly procedure. The results now report the workflow used for the analysis, and this should be specified along with the BUSCO dataset when reporting scores in publications.

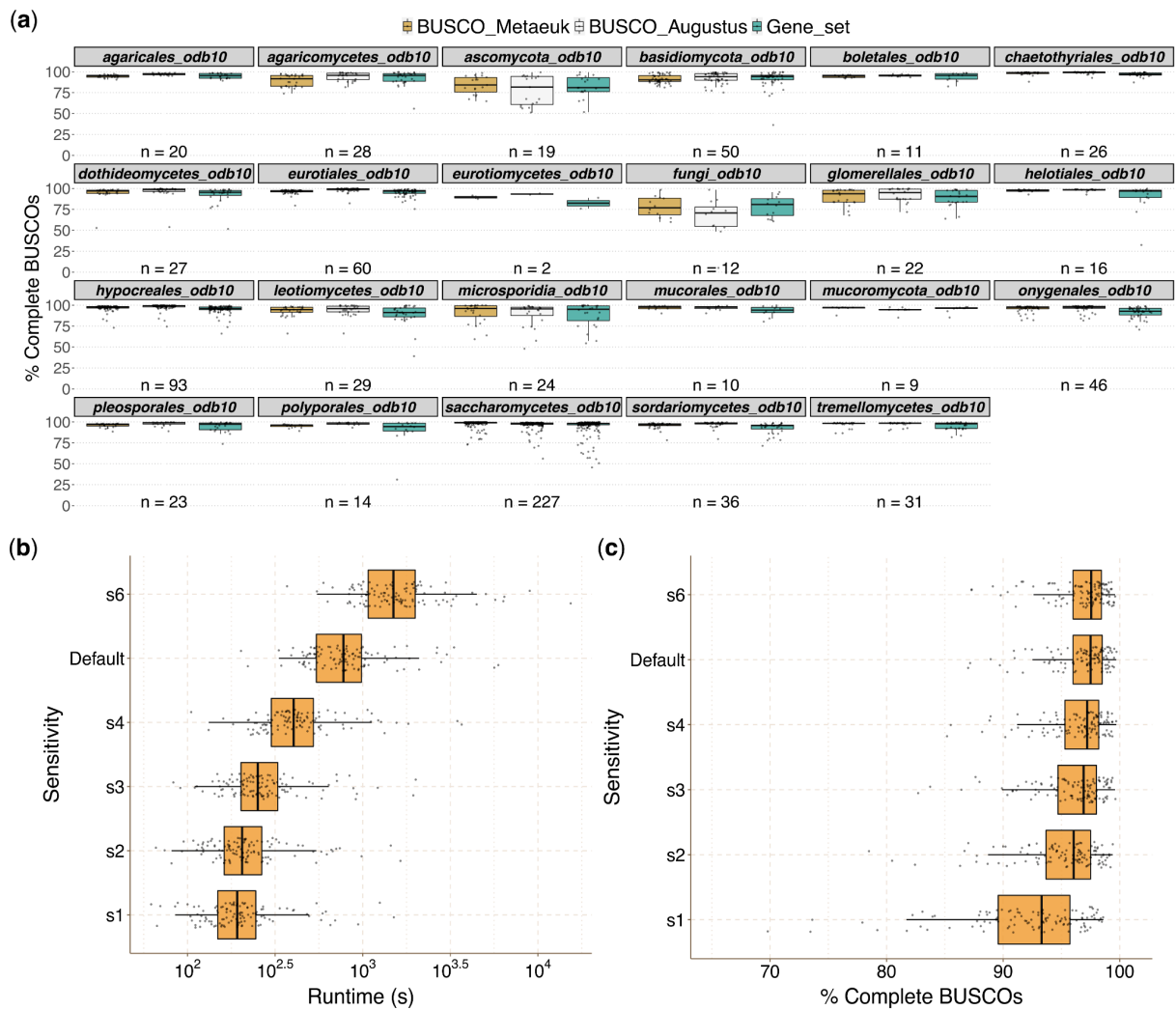

**Fig. 2.** (*a*) Comparisons of BUSCO scores obtained on a set of fungal genomes using the two available workflows for eukaryotic species. The percentage on the y-axis corresponds to the complete BUSCOs for the BUSCO_MetaEuk (orange) and BUSCO_Augustus (white) workflows. Assessments on gene sets are also displayed for comparison (green). Genomes were assessed using the most specific available datasets, which are displayed at the top of each sub-panel. The newly-introduced BUSCO_MetaEuk workflow allows faster assessments, see supplementary figure 3a for the differences in runtimes. (*b* and *c*) Effect of using different MetaEuk sensitivity values on BUSCO_Metaeuk runtimes and completeness estimation for 112 arthropod genomes evaluated with their most specific BUSCO dataset. The default values are set at s=4.5 and s=6 for the first and the second MetaEuk runs, respectively. For the analyses, the same sensitivity value displayed on the y-axis was used for both MetaEuk runs. The axis corresponding to runtimes (in seconds) is log-transformed.

***Auto-lineage workflow: an all-in-one option for quality assessment of (meta)genomic data.***
BUSCO v5 is the only available tool that can assess genomic data from the three domains of life in a single analysis by using the new "--auto-lineage" function. This is achieved through the phylogenetic placement of the input sequence (genome, gene set or transcriptome) on a set of pre-computed phylogenetic trees using SEPP (Mirarab et al. 2011) and pplacer (Matsen et al. 2010). Subsequently, BUSCO automatically attempts to select the most specific (i.e. highest-resolution) dataset available for the species of interest. This solves a major problem when analyzing metagenomic data where the taxonomic origin of MAGs is often unknown and both eukaryotic and prokaryotic genomes can be present in the sample. For ease of batch analyses, it is now possible to run BUSCO on multiple input sequences, and an additional table summarizing the scores for all inputs is returned. Additionally, BUSCO can automatically detect a subset of viruses belonging to clades of the 27 newly-introduced viral datasets. Supplementary figure 5 and supplementary table 6 show the assessment on RefSeq (Brister et al. 2015) viral genomes and gene sets using the "--auto-lineage" function. We plan to expand the virus pipeline in future to allow assessments of a broader set of viruses. An overview of the auto-lineage workflow can be found in the supplementary notes.

Figure 3a presents the performance, in terms of selecting the right dataset, of the auto-lineage procedure for 436 bacterial/archaeal genomes (supplementary table 7). We compared BUSCO v5 with CheckM v1.1.3 (Parks et al. 2015), which, while conceptually similar, is only capable of performing assessments on bacterial and archaeal data. Figure 3b and c reports BUSCO and CheckM completeness and redundancy scores for the same set of 436 genomes (see also supplementary table 8). BUSCO estimates appear to be more conservative for some data points. This is in part related to the higher resolution datasets (i.e. more markers) automatically selected for the assessment (supplementary fig. S6). Overall, on prokaryotic data, BUSCO has comparable results to CheckM. However, a major advantage of BUSCO is the ability to detect and assess eukaryotic microbial genomes.

In terms of resources, the mean runtime per genome with BUSCO (on the 436 prokaryotic genomes, in batch mode and using the prokaryotic-specific "--auto-lineage-prok") is less than a minute (e.g. 35s and 48s with 30 and 8 CPUs, respectively) (supplementary table 9, 10). Running BUSCO through a workflow management system can considerably reduce the overall runtime. We provide an example using Snakemake (Mölder et al. 2021) at https://gitlab.com/ezlab/plugins_buscov5. With this setup and allowing a total number of 30 CPUs with 5 CPUs per task, the overall runtime for completing the same assessment was reduced to 95 min (a mean runtime of 13s per genome) compared to 257 min. Figure 3d shows the memory requirements for assessing a set of bacterial and fungal genomes with the BUSCO auto-lineage workflow. The memory requirements do not exceed 11GB for bacterial genomes, so that they can be assessed on laptops with limited memory and CPU resources (e.g. the same assessment ran to completion on a MacBook Pro with 16GB and 8 cores). This is an advantage in comparison to the 70 GB of memory required by CheckM (supplementary table 9).

The percentage of duplicated markers reported by BUSCO can reflect technical or biological duplications (i.e. redundant markers derive from the same genome), or contaminations from other species/strains. BUSCO cannot directly distinguish whether redundant BUSCOs are due to duplications or contamination. Nevertheless the sequences detected by BUSCO as "duplicated" can be used as evidence for further investigation. In general, a high "duplication" score for prokaryotic data is more likely to be caused by contamination rather than gene duplication, especially for MAGs, which often require manual refinements (Eren et al. 2015). Figure 3c shows an overall concordance between BUSCO "duplication" and CheckM "contamination" estimates.

Additionally, BUSCO v5 can highlight contamination deriving from species belonging to other domains by means of the scores obtained from assessing the input with the three "root" datasets (bacteria_odb10, archaea_odb10, eukaryota_odb10). These dataset assessments are automatically performed as the first step of the "auto-lineage" workflow. As there can be a background level of cross-matches between datasets, it is expected to have BUSCOs scoring in multiple "root" datasets. For example, running the "root" datasets on a set of 2'779 bacterial genomes from RefSeq (O'Leary et al. 2016) results in a median complete BUSCO score of 18.5% and 4.3% for the archaea_odb10 and eukaryota_odb10 datasets, respectively (supplementary table 10). The frequency of matches (including those reported as "fragmented") for each BUSCO is shown in supplementary figure 7. 214 (83.9%) of the eukaryota_odb10 BUSCOs are detected in 5% or fewer of the bacterial genomes, with 184 (72.2%) with no matches at all. 120 (61.9%) of the archaea_odb10 BUSCOs are detected in 5% or fewer of the bacterial genomes, with 87 (44.84%) never showing up (supplementary table 11). The percentage of complete matches from bacteria_odb10 and archaea_odb10 in eukaryotic genomes is, on average, higher, e.g. 12.1% and 21.1% in fungal genomes and of 30.7% and 62.4% in arthropod gene sets, respectively (supplementary fig. 8, supplementary table 10 and supplementary table 11), making it less useful to spot contaminants. Nevertheless, high completeness scores from these datasets may alert users to the potential presence of contaminant species or horizontal gene transfer events from other domains, whereas high duplication scores may indicate the presence of multiple contaminant species.

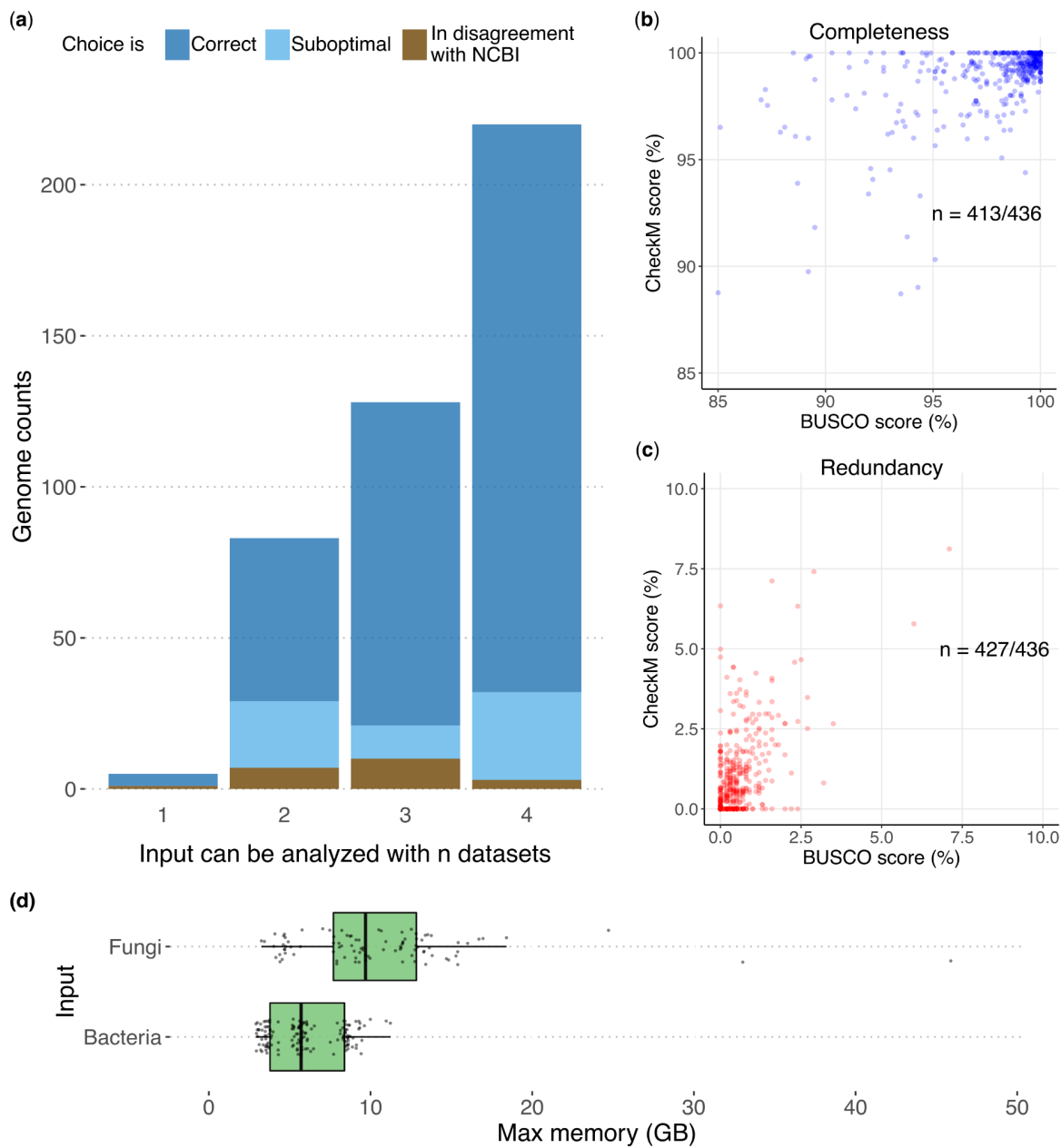

**Fig. 3.** BUSCO assessment on microbial data and comparison with CheckM. (*a*) Accuracy in the choice of dataset produced by the auto-lineage mode when analyzing bacterial and archaeal assemblies (n=436). For a given assembly, there can be between one and four suitable datasets (from the more general, root dataset, down to the more specific one) to choose from (x-axis). The selected dataset is considered as "correct" when it is the most lineage-specific available for the genome; "suboptimal" when a parent lineage is selected; and "in disagreement with the NCBI" when the selected lineage is not part of the NCBI taxonomic annotation of that genome. This might indicate an error; however, 12 out of 19 genomes in this category are annotated by NCBI as "unclassified", while sharing a parent lineage with the BUSCO selected dataset; e.g. assembly GCF_000153385.1 is an unclassified Flavobacteria and was assigned to flavobacteriales_odb10 dataset (also see supplementary table 7). When supported by a high BUSCO score, this suggests that the dataset selected by BUSCO was appropriate. (*b* and *c*) Comparison of BUSCO and CheckM completeness (blue) and redundancy (red)

scores on a set of 436 genomes. For clarity, the two scatterplots are zoomed in on the areas of highest densities. n represents the number of data points displayed in the zoomed area. (*d*) Memory requirements for running BUSCO with the auto-lineage workflow on a set of bacterial and fungal genomes.

*Benchmark of predictions*

To assess the precision of the BUSCO estimates, we benchmarked BUSCO v5 predictions on gene sets and genomes artificially depleted of randomly selected genes. Briefly, we randomly removed 0/10/30/50% of the genes in the official gene set, generating five simulated versions for each level of depletion. The corresponding genes in the genome were masked using the coordinates from the GFF file. BUSCO v5 was run on the simulated gene sets and genomes (both BUSCO_Augustus and BUSCO_MetaEuk workflows) using the most specific dataset and the most generic one (for more details see supplementary notes). Figure 4a shows an example of the results of the benchmarks for *Drosophila melanogaster* (assembly accession: GCF_000001215.4). The predicted BUSCO estimates on depleted gene sets and genomes have a good overall correspondence to the expected values when using the diptera_odb10 dataset, the most specific and appropriate dataset. When using the more generic eukaryota_odb10 dataset, the estimates are subjected to more variability (supplementary fig. 9a) across the different depleted versions, which is expected and explained by the lower number of markers, and the correspondingly lower coverage of the genome/gene set. This further highlights the importance of using the most specific dataset when possible.

A second slightly different type of benchmarking was used to compute the number of false positive (FP)/false negative (FN) predictions. In this case we exclusively depleted genes predicted to be BUSCO markers. We first mapped the gene set to the OrthoDB level of interest (e.g. if assessing the species with diptera_odb10, the gene set was mapped to the OrthoDB diptera level). Based on this ortholog mapping we depleted the gene set/assemblies by 0/10/30/50/100% of the predicted BUSCO genes, generating five versions for each depletion level (except for 100% depletion). BUSCO v5 was run on these simulated data and the FP/FN and precision estimates were computed using the initial orthoDB mapping as ground truth. Figure 4b shows the overall congruent estimates on *D. melanogaster* data. Removing 100% of the potential markers results in a small percentage of false positives, with the newly introduced BUSCO_MetaEuk workflow having a smaller number of false positives. All three modes have similar precision (fig. 4c), with BUSCO_MetaEuk workflow showing a slightly higher precision when depletion is equal to or above 50%. Supplementary figure 10 shows the results of the two benchmarks on the yeast *Saccharomyces cerevisiae* (GCF_000146045.2). A detailed description of the benchmark procedures is reported in the supplementary notes.

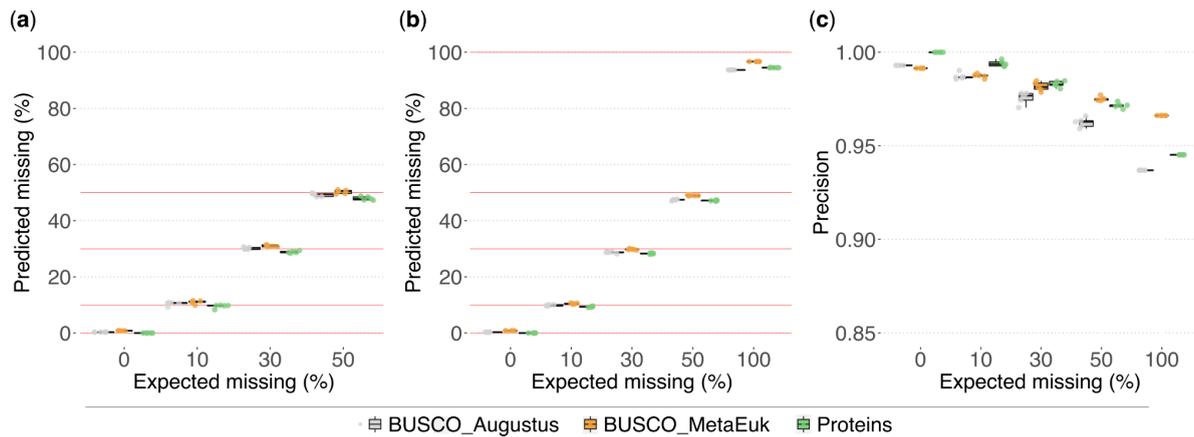

**Fig. 4**. Benchmarking BUSCO estimates on artificially depleted genomes and gene sets of *Drosophila melanogaster* assessed with the diptera_odb10 dataset. (*a*) Artificial depletion was made on the full gene set. (*b*) Artificial depletion exclusively made on genes matching BUSCO markers. For both panels, solid red lines indicate the expected missing values. Five randomly depleted versions were used for each level of depletion. (*c*) Precisions of the predictions for the analyses of panel b.

*Other improvements and distribution*

Along with a major refactoring of the code, new options have been added for managing the increasing number of datasets and to facilitate the analysis procedure. In particular, we added a default option to automatically download the necessary precomputed files for phylogenetic placement, and the datasets, either by specifying "-l <dataset_name>" as an option on the command line when initiating a BUSCO run, or by running in auto-lineage mode. Documentation and software setup instructions are all described in detail at https://busco.ezlab.org/busco_userguide.html. We now also maintain a BUSCO package on Bioconda (https://anaconda.org/bioconda/busco) and a Docker container. We encourage BUSCO users to favor these two approaches to control the version of each software dependency that is used. In addition, the BUSCO code is still distributed on GitLab https://gitlab.com/ezlab/busco.

## Materials and Methods

BUSCO datasets are available at https://busco-data.ezlab.org/v5/data/lineages/. Each BUSCO dataset contains the details on the species, orthologous groups and genes used to construct the set. Versions and accessions of all the genome assemblies and gene sets, and the BUSCO main results analyzed as part of this study are listed in the supplementary tables. Further details on the analyses are described in the supplementary notes. Plots presenting the results of the analyses were made using the ggplot2 package (Wickham 2009) in R (R Core Team 2020).

## Acknowledgment

This work was supported by funding from the University of Geneva, Swiss National Science Foundation grant 310030_189062 and Swiss Institute of Bioinformatics SERI to EZ. We would like to thank all current and former members of the EZ group, and Robert M. Waterhouse for continuing support of the BUSCO project and the user community.

## Data Availability

BUSCO is licensed and freely distributed under the MIT Licence. The BUSCO source code is available through the GitLab project, https://gitlab.com/ezlab/busco, and it is also maintained on Bioconda (https://anaconda.org/bioconda/busco) and as a Docker container. BUSCO datasets are available at https://busco-data.ezlab.org/v5/data/lineages/. Each BUSCO dataset contains the details on the species, orthologous groups and genes used to construct the set. Further information can be accessed by mapping through the OrthoDB website. We also provide a single text file at https://busco-data.ezlab.org/v5/data/ reporting the IDs of all proteins used to build the datasets with all the mappings to relevant information (e.g. dataset, Orthologous Group, species name, assemblyID). Versions and accessions of all the genome assemblies and gene sets analyzed in this study, along with their corresponding species name are listed in the supplementary tables, and are accessible through the NCBI database. The simulated data and the intermediate outputs underlying the benchmark of predictions are available on Zenodo, at https://www.doi.org/10.5281/zenodo.4972052. Supplementary tables and notes are available at https://gitlab.com/ezlab/busco_preprint_2021.

# References

Brister JR, Ako-adjei D, Bao Y, Blinkova O. 2015. NCBI Viral Genomes Resource. *Nucleic Acids Research* 43:D571.

Eren AM, Esen ÖC, Quince C, Vineis JH, Morrison HG, Sogin ML, Delmont TO. 2015. Anvi'o: an advanced analysis and visualization platform for 'omics data. *PeerJ* 3:e1319.

Grüning B, Dale R, Sjödin A, Chapman BA, Rowe J, Tomkins-Tinch CH, Valieris R, Köster J. 2018. Bioconda: sustainable and comprehensive software distribution for the life sciences. *Nature Methods* 15:475–476.

Hyatt D, Chen G-L, LoCascio PF, Land ML, Larimer FW, Hauser LJ. 2010. Prodigal: prokaryotic gene recognition and translation initiation site identification. *BMC Bioinformatics* 11:119.

Kriventseva EV, Kuznetsov D, Tegenfeldt F, Manni M, Dias R, Simão FA, Zdobnov EM. 2019. OrthoDB v10: sampling the diversity of animal, plant, fungal, protist, bacterial and viral genomes for evolutionary and functional annotations of orthologs. *Nucleic Acids Research* 47:D807–D811.

Levy Karin E, Mirdita M, Söding J. 2020. MetaEuk—sensitive, high-throughput gene discovery, and annotation for large-scale eukaryotic metagenomics. *Microbiome* 8:48.

Matsen FA, Kodner RB, Armbrust EV. 2010. pplacer: linear time maximum-likelihood and Bayesian phylogenetic placement of sequences onto a fixed reference tree. *BMC Bioinformatics* 11:538.

Merkel D. 2014. Docker: lightweight Linux containers for consistent development and deployment. *Linux J.* 2014:2:2.

Mirarab S, Nguyen N, Warnow T. 2011. SEPP: SATé-Enabled Phylogenetic Placement. In: Biocomputing 2012. Kohala Coast, Hawaii, USA: WORLD SCIENTIFIC. p. 247–258. Available from: http://www.worldscientific.com/doi/abs/10.1142/9789814366496_0024

Mölder F, Jablonski KP, Letcher B, Hall MB, Tomkins-Tinch CH, Sochat V, Forster J, Lee S, Twardziok SO, Kanitz A, et al. 2021. Sustainable data analysis with Snakemake. *F1000Res* 10:33.

Nayfach S, Camargo AP, Schulz F, Eloe-Fadrosh E, Roux S, Kyrpides NC. 2020. CheckV assesses the quality and completeness of metagenome-assembled viral genomes. *Nat Biotechnol.*

O'Leary NA, Wright MW, Brister JR, Ciufo S, Haddad D, McVeigh R, Rajput B, Robbertse B, Smith-White B, Ako-Adjei D, et al. 2016. Reference sequence (RefSeq) database at NCBI: current status, taxonomic expansion, and functional annotation. *Nucleic Acids Research* 44:D733–D745.

Parks DH, Imelfort M, Skennerton CT, Hugenholtz P, Tyson GW. 2015. CheckM: assessing the quality of microbial genomes recovered from isolates, single cells, and metagenomes. *Genome Res.* 25:1043–1055.

Parra G, Bradnam K, Korf I. 2007. CEGMA: a pipeline to accurately annotate core genes in eukaryotic genomes. *Bioinformatics* 23:1061–1067.

R Core Team. 2020. R: A Language and Environment for Statistical Computing. Vienna, Austria: R Foundation for Statistical Computing Available from: https://www.R-project.org/

Saary P, Mitchell AL, Finn RD. 2020. Estimating the quality of eukaryotic genomes recovered from metagenomic analysis with EukCC. *Genome Biology* 21:244.

Simão FA, Waterhouse RM, Ioannidis P, Kriventseva EV, Zdobnov EM. 2015. BUSCO: assessing genome assembly and annotation completeness with single-copy orthologs. *Bioinformatics* 31:3210–3212.

Stanke M, Diekhans M, Baertsch R, Haussler D. 2008. Using native and syntenically mapped cDNA alignments to improve de novo gene finding. *Bioinformatics* 24:637–644.

Steinegger M, Söding J. 2017. MMseqs2 enables sensitive protein sequence searching for the analysis of massive data sets. *Nature Biotechnology* 35:1026–1028.

Wadi L, Reinke AW. 2020. Evolution of microsporidia: An extremely successful group of eukaryotic intracellular parasites. *PLOS Pathogens* 16:e1008276.

Waterhouse RM, Seppey M, Simão FA, Manni M, Ioannidis P, Klioutchnikov G, Kriventseva EV, Zdobnov EM. 2018. BUSCO Applications from Quality Assessments to Gene Prediction and Phylogenomics. *Mol Biol Evol* 35:543–548.

Wickham H. 2009. ggplot2: Elegant Graphics for Data Analysis. New York: Springer-Verlag Available from: https://www.springer.com/de/book/9780387981413

Zdobnov EM, Kuznetsov D, Tegenfeldt F, Manni M, Berkeley M, Kriventseva EV. 2021. OrthoDB in 2020: evolutionary and functional annotations of orthologs. *Nucleic Acids Research* 49:D389–D393.

## Tables

**Table 1:** Number of odb9 and odb10 BUSCO datasets. The odb10 version greatly expanded the number of benchmarking datasets.

| Dataset | odb9 (v3) | odb10 (v4/5) |
|---|---|---|
| Bacteria | 16 | 83 |
| Archaea | 0 | 16 |
| Viruses | 0 | 27 |
| Eukaryota | 33 | 67 |
|   Protist | 2 | 7 |
|   Fungi | 10 | 24 |
|   Plants | 1 | 9 |
|   Metazoa | 14 | 26 |
|     Arthropoda | 5 | 8 |
|     Vertebrata | 7 | 15 |
| Total | 49 | 193 |

## Legends of Supplementary Tables

**Supplementary table 1.** Datasets available for BUSCO version 4 and 5 (*_odb10) along with the number of BUSCO markers and the number of species used to build the datasets. Corresponding values are shown for equivalent datasets available in version 3 (*_odb9). (this table is in the supplementary notes pdf file)

**Supplementary table 2.** Table of BUSCO scores obtained by running BUSCO v3 and v5 on bacterial, fungi and metazoa gene sets using the bacteria_odb9/odb10, fungi_odb9/odb10 and metazoa_odb9/odb10 datasets respectively.

**Supplementary table 3.** Breakdown of the main differences among BUSCO v3, v4 and v5. (this table is in the supplementary notes pdf file)

**Supplementary table 4.** Table of scores obtained by running the BUSCO_MetaEuk and BUSCO_Augustus workflows on fungi, protist and arthropod genomes.

**Supplementary table 5.** Table of scores and runtimes obtained by running BUSCO_MetaEuk workflows using different MetaEuk sensitivity values.

**Supplementary table 6.** Table of scores obtained by running BUSCO on a set of viral genomes and gene sets from RefSeq for which a BUSCO viral dataset is available.

**Supplementary table 7.** BUSCO auto-lineage dataset selection on 436 bacterial and archaeal genomes.

**Supplementary table 8.** Table of scores obtained by running BUSCO (auto-lineage-prok) and CheckM batch mode (lineage_wf) on 436 bacterial and archaeal genomes.

**Supplementary table 9.** Summary table of the runtimes and memory requirements for running BUSCO (auto-lineage-prok) and CheckM batch mode (lineage_wf) on 436 bacterial and archaeal genomes using 30, 12 and 8 CPUs. The memory and runtime values were recorded using the "benchmark" feature of Snakemake. (this table is in the supplementary notes pdf file)

**Supplementary table 10.** Runtimes and memory requirements for each BUSCO assessment of the 436 bacterial and archaeal genomes using 30, 12 and 8 CPUs. The values were recorded using the "benchmark" feature of Snakemake.

**Supplementary table 11.** Scores from cross-domain matches on bacterial genomes (with archaea_odb10 and eukaryota_odb10), fungal genomes and invertebrates gene sets (with bacteria_odb10, archaea_odb10).

**Supplementary table 12.** Frequency of cross-domain matches for each BUSCO marker from the eukaryota_odb10 and archaea_odb10 datasets in 2'779 RefSeq bacterial genomes, and from the bacteria_odb10 and archaea_odb10 datasets in 370 fungal genomes and 235 invertebrates gene sets from RefSeq.